\documentstyle[aps,prl,epsf,preprint]{revtex}

\newcommand{\hD}{\hat D}

\begin{document}


\preprint{YITP-00-1, gr-qc/0001015}

\draft

\begin{titlepage}
\title{
  Dynamics of a string coupled to gravitational waves\\
  \smallskip 
  {\normalsize \it
     --- Gravitational wave scattering by a Nambu-Goto straight string ---
   }
}
\author{
  Kouji NAKAMURA$^{1}$\footnote{E-mail:kouchan@phys-h.keio.ac.jp}, 
  Akihiro ISHIBASHI$^{2}$\footnote{E-mail:akihiro@yukawa.kyoto-u.ac.jp}
  and    
  Hideki ISHIHARA$^{3}$\footnote{E-mail:ishihara@th.phys.titech.ac.jp}
}
\address{
  ${}^{1}$Department of Physics, Keio University,
  Hiyoshi Yokohama, 223-8521, Japan
}
\address{
  ${}^{2}$Yukawa Institute for Theoretical Physics, Kyoto
  University, Kyoto 606-8502, Japan
}
\address{
  ${}^{3}$Tokyo Institute of Technology,
  Oh-Okayama Meguro-ku, Tokyo 152-0033, Japan
}
\date{January 5, 2000}
\maketitle

\begin{abstract}
We study the perturbative dynamics of an infinite gravitating 
Nambu-Goto string within the general-relativistic perturbation
framework.
We develop the gauge invariant metric perturbation on a
spacetime containing a self-gravitating straight string with a
finite thickness and solve the linearized Einstein equation.
In the thin string case, we show that the string does not emit
gravitational waves by its free oscillation in the first order
with respect to its oscillation amplitude, nevertheless the
string actually bends when the incidental gravitational waves go
through it.
\end{abstract}
\end{titlepage}



There are significant interests in topological defects formed
during phase transition in the early
universe\cite{Vilenkin_Shellard,Kolb_Turner}.  
In particular, it has been thought that these defects radiate
gravitational wave by their rapid oscillation\cite{Vachaspati}.
Thus, it is crucially important to study the precise dynamics of
the defects and their gravitational effects.


In the simplest case, the defects are idealized by the
infinitesimally thin Nambu-Goto membranes, i.e., their dynamics
is governed by the minimization of their world hyper-volume. 
If the self-gravity of membranes is ignored (test membrane
case), the Nambu-Goto action admits oscillatory solutions. 
It is considered that the 
membranes gradually lose their kinetic energy by the
gravitational wave emission\cite{Vachaspati}.  
However, by taking into account the self-gravity of the Nambu-Goto
wall, it is shown that a self-gravitating wall coupled to
gravitational wave behaves in a quite different manner\cite{Kodama}. 
The dynamical degree of freedom concerning the perturbative
oscillations around a spherical one is given by that of gravitational
waves and self-gravitating spherical walls do not oscillate
spontaneously unlike test walls.


How about Nambu-Goto strings?
It is also considered that the cosmic strings oscillate rapidly
and gradually lose their kinetic energy by the gravitational
wave emission\cite{Vachaspati}.
The energy momentum tensor of an oscillating infinite test
string and the gravitational wave emission are studied by
several authors\cite{Sakellariadou,traveling_wave-Vachaspati}.
However, it is not clear how a Nambu-Goto string behaves when
one takes into account of its self-gravity.



In this paper, we consider the perturbative oscillation of
an infinite self-gravitating string using a exactly solvable
model within the general-relativistic perturbation framework and
show that a self-gravitating infinite string behaves in the same
manner as the above self-gravitating wall in the first order
with respect to the oscillation amplitude of the string.


It is known that the mathematical description of a thin string
is more delicate than a domain wall because the support of a
string is a surface of co-dimension two.
There is no simple prescription of an arbitrary line source
where a metric becomes singular\cite{Geroch}.
In this paper, we consider, first, a straight string with a
finite thickness so that the singularity is
regularized\cite{Vilenkin_Shellard,stright_string}.  
Then, the metric junction formalism is applicable on the surface
of the thick string.  
Next, we consider gravitational wave emission by the thick
string motion which is excited by incident gravitational wave,
i.e., the scattering problem of gravitational wave by a thick
string. 
We analyze this problem by the gauge invariant linear
perturbation theory and show the perturbative velocity of the
string is given by the variable of gravitational waves.
In the thin string case, we show that there is neither resonance
nor phase shift in the gravitational wave scattered by a
Nambu-Goto string, nevertheless the string is bent by
gravitational waves. This shows that self-gravitating infinite
strings do not oscillate spontaneously unlike test string at
least in the first order with respect to its oscillation
amplitude.



As the background for the perturbation, we consider a spacetime
$({\cal M},g_{\mu \nu})$ containing a straight thick string.
The surface ${\cal S}$ of the thick string divides ${\cal M}$
into two regions: $ {\cal M}_{ex}$ and ${\cal M}_{in}$. 
Note that ${\cal M}_{in}$ describes the `thick' worldsheet of 
the string. 
We assume that the spacetime ${\cal M}$ is static and
cylindrically symmetric. 
Then we divide ${\cal M}$ into two submanifolds so that 
${\cal M} = {\cal M}_{1}\times{\cal M}_{2}$ and write the
background metric on ${\cal M}$ in the form
\begin{equation}
  \label{bg_metric}
  ds^{2} = \gamma_{ab}dy^{a}dy^{b} + \eta_{pq}dz^{p}dz^{q},
\end{equation}
where $\gamma_{ab}$, the metric on ${\cal M}_{1}$, and
$\eta_{pq}$, that on ${\cal M}_{2}$, are given by 
\begin{equation}
  \label{bg_metric-2}
  \gamma_{ab}dy^{a}dy^{b} = d\rho^{2} + r(\rho)^{2} d\phi^{2}, \quad
  \eta_{pq}dz^{p}dz^{q} = - dt^{2} + dz^{2}, \quad 0\leq\phi\leq 2\pi.
\end{equation} 
We shall use the indices $a,...,d$ for tensors on ${\cal
M}_{1}$ and $p,...,s$ for those on ${\cal M}_{2}$. 
The string thickness is given by the circumference radius
$r_{*}$ of ${\cal S}$.


Since a Nambu-Goto string is characterized by a constant
string tension $\sigma_0$, we consider the following
energy-momentum tensor,
\begin{equation}
  \label{ene_mon}
  T_{\mu\nu} = - \sigma \eta_{\mu\nu},
\end{equation}
where $ \sigma = \sigma_0 $ for $r <r_{*}$ and $\sigma = 0$ for
$ r\ge r_*$ \cite{Vilenkin_Shellard}.  
Here $\eta_{\mu\nu}$ is the four dimensional extension of
$\eta_{pq}$.

\begin{figure}[htbp]
  \begin{center}
    \leavevmode
    \epsfxsize=8cm
    \epsfbox[0 0 620 512]{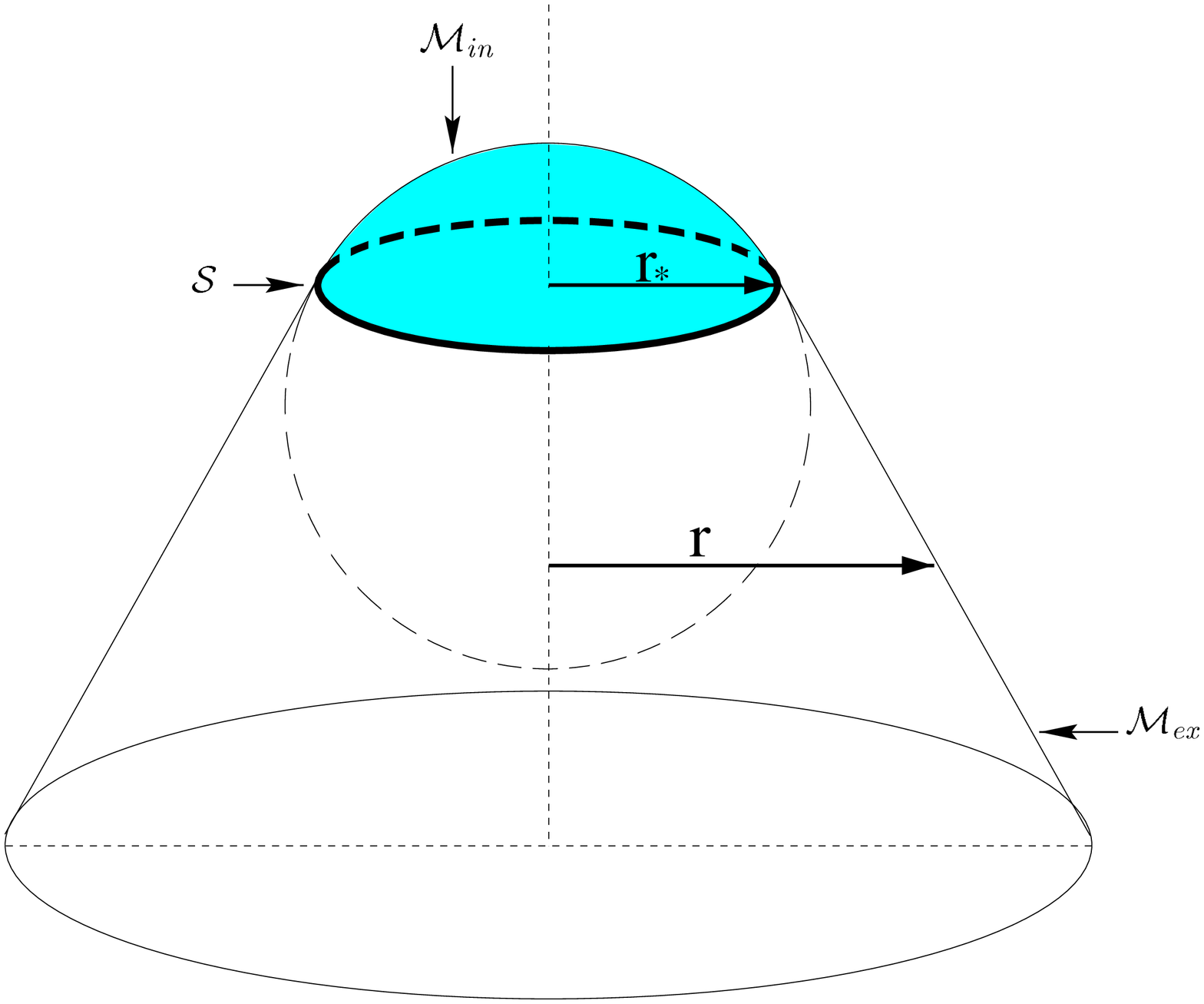}
    \caption{The global geometry of  ${\cal M}_{1}$.}
    \label{fig:fig1}
  \end{center}
\end{figure}

The Einstein equations for the metric (\ref{bg_metric}) and
(\ref{bg_metric-2}) are reduced to the single equation
\begin{equation}
  {\cal R} = - 2 \frac{\partial_{\rho}^{2}r}{r} = 16\pi G \sigma,
\end{equation}
where ${\cal R}$ is the Ricci curvature on ${\cal M}_{1}$. 
The solution on ${\cal M}_{in}\cap{\cal M}_{1}$ is given by 
\begin{equation}
  \label{homo_BG}
  \gamma_{ab}dy^{a}dy^{b} 
  = \frac{dr^{2}}{1 - \hat{\alpha}^{2}r^{2}} + r^{2}d\phi^{2}, 
  \quad \hat{\alpha}^{2} =  \frac{{\cal R}}{2},
\end{equation}
and that on ${\cal M}_{ex}\cap{\cal M}_{1}$ is 
\begin{equation}
  \label{vacuum_BG}
  \gamma_{ab}dy^{a}dy^{b} =  \frac{dr^{2}}{(1 - \alpha)^{2}} +
  r^{2}d\phi^{2}, 
\end{equation}
where $\alpha$ is a deficit angle on ${\cal M}_{ex}$. 
These two solutions are joined along the surface ${\cal S}$ by
Israel's junction condition\cite{Israel}:
$[K^{\mu}_{\;\;\nu}] := K^{\mu}_{\;\;\nu+} - K^{\mu}_{\;\;\nu-} = 0,$ 
where $K^{\mu}_{\;\;\nu\pm}$ are the extrinsic curvature of
${\cal S}$ facing to ${\cal M}_{ex}$ and ${\cal M}_{in}$,
respectively.  
For the solutions (\ref{homo_BG}) and (\ref{vacuum_BG}), this
junction condition is reduced to 
$\alpha = 1 - \sqrt{1 - \hat{\alpha}^{2} r_{*}^{2}}$.
The global geometry of ${\cal M}_{1}$ is illustrated in
Fig.\ref{fig:fig1}.



%
%
%
%
%
We consider the metric perturbations on the background geometry
given by (\ref{bg_metric}), (\ref{homo_BG}) and (\ref{vacuum_BG}). 
Let $h_{\mu\nu}$ be a perturbative metric and
$t^{\mu}_{\;\;\nu}$ be a perturbed energy-momentum tensor, which
can be expanded by the harmonics on ${\cal M}_{2}$ as follows: 
\begin{eqnarray}
  \label{metric_decomp_def_1}
  &h_{ab} = {\displaystyle \int} f_{ab}S, \quad
  h_{ap} = {\displaystyle \int} \left\{f_{a(o1)}
  V_{(o1)p} + f_{a(e1)} V_{(e1)p}\right\},& \\
  \label{metric_decomp_def_2}
  &h_{pq} = {\displaystyle \int} \left\{f_{(o2)}
    T_{(o2)pq} + f_{(e0)} T_{(e0)pq}  + f_{(e2)} T_{(e2)pq}\right\},& \\
  \label{e-m_decomp_def_1}
  &t^{a}_{\;\;b} = {\displaystyle \int} s^{a}_{\;\;b}S, \quad
  t^{a}_{\;\;p} = {\displaystyle \int} \left\{ s^{a}_{(o1)} V_{(o1)p} 
    + s^{a}_{(e1)} V_{(e1)p}\right\},& \\
  \label{e-m_decomp_def_2}
  &t^{p}_{\;\;q} = {\displaystyle \int} \left\{s_{(o2)}
    {T_{(o2)}}^{p}_{\;\;q} + s_{(e0)} {T_{(e0)}}^{p}_{\;\;q} 
    + s_{(e2)} {T_{(e2)}}^{p}_{\;\;q}\right\}.& 
\end{eqnarray}
Here $\int := \int d\omega dk_{z}$ and 
\begin{eqnarray}
  S := e^{-i\omega t + i k_{z}z}, \quad &V_{(o1)}^{p} :=
  \epsilon^{pq}\hD_{q}S,& \quad V_{(e1)}^{p} :=
  \eta^{pq}\hD_{q}S, \nonumber\\ 
  T_{(e0)pq} := \frac{1}{2}\eta_{pq} S, \quad &T_{(e2)pq} :=
  \left(\hD_{p}\hD_{q} - \frac{1}{2}\eta_{pq}
    \hD^{r}\hD_{r} \right)S,& \quad T_{(o2)pq} := -
  \epsilon_{r(p}\hD_{q)}\hD^{r} S, 
  \label{harmonics-kneq0}
\end{eqnarray}
are independent tensor harmonics on ${\cal M}_{2}$, 
$\epsilon_{rs}$ is a two-dimensional antisymmetric tensor on
${\cal M}_{2}$, and $\hat{D}_p$ denotes the covariant derivative
associated with $\eta_{pq}$.  
The symbols $(o)$ and $(e)$ refer to odd and even parity modes
with respect to the inversion of $(t,z)$, respectively.  
The expansion coefficients are tensors on ${\cal M}_{1}$.
The perturbative energy momentum tensor $t^{\mu}_{\;\;\nu}$ has
its support only on ${\cal M}_{in}$.


We define $\kappa^{2} := \omega^{2} - k_{z}^{2}$, which is the eigen
value of a differential operator $\eta^{pq}\hD_{p}\hD_{q}$.
We note that the mode with $\kappa=0$, which propagates along
the string, is not included in the expansion
(\ref{metric_decomp_def_1})-(\ref{e-m_decomp_def_2}). 
The $\kappa=0$ mode will be discussed in Ref.\cite{kouchan-k=0}.


Here we consider the gauge-transformation of $h_{\mu\nu}$ and
$t^{\mu}_{\;\;\nu}$ associated with 
$x^{\mu} \rightarrow x^{\mu} + \xi^{\mu}$, where $\xi^{\mu}$ is
expanded as 
\begin{equation}
  \label{xi_decomp_def}
  \xi_{a} := {\displaystyle \int} \zeta_{a}S, \quad
  \xi_{p} := {\displaystyle \int} \left\{\zeta_{(o1)} V_{(o1)p} 
    + \zeta_{(e1)} V_{(e1)p}\right\}. 
\end{equation}
Inspecting the gauge transformed variables 
$h_{\mu\nu} - {\pounds}_{\xi}g_{\mu\nu}$ and 
$t^{\mu}_{\;\;\nu} - {\pounds}_{\xi}T^{\mu}_{\;\;\nu}$, we find
simple gauge-invariant combinations of the expansion
coefficients: for odd modes,  
\begin{equation}
  \label{F_a_def}
  F_{a} := f_{a(o1)} - \frac{1}{2}D_{a}f_{(o2)}, 
\end{equation}
and for even modes, 
\begin{eqnarray}
  \label{F_ab_F_def}
  && F_{ab} := f_{ab} - D_{a}X_{b} - D_{b}X_{a}, \quad
  F := f_{(e0)} - \kappa^2 f_{(e2)},
\end{eqnarray}
where $D_a$ is a covariant derivative associated with
$\gamma_{ab}$ and the variable 
$X^{a} := f^{a}_{(e1)} - \frac{1}{2} D^{a}f_{(e2)}$ is
transformed to $X^{a} - \zeta^{a}$ by the gauge transformation.


We also introduce the gauge invariant variables for
the perturbations of $T^{\mu}_{\;\;\nu}$ by 
\begin{equation}
  \Sigma := 16\pi G (s_{(e0)} + 2 X^{a}D_{a}\sigma), \quad
  V^{a} := 16\pi G (s^{a}_{(e1)} - \sigma X^{a}). 
\end{equation}
Note that all the expansion coefficients except for $s_{(e0)}$
and $s^{a}_{(e1)}$  are gauge invariant by themselves. 
In this article, we consider the perturbative motion of a
Nambu-Goto string in the first order with respect to its
oscillation amplitude. Within this order, $\Sigma$ is the energy
density perturbation which is equal to the tangential tension of
a string and $V^{a}$ corresponds to the momentum perturbation. 
The other coefficients, within the same order, $s^{a}_{\;\;b}$,
$s^{a}_{(o1)}$, $s_{(o2)}$ and $s_{(e2)}$ are regarded as the
tension normal to the string worldsheet, spin of the string,
Lorentz boost along the string and the energy density
perturbation which is not equal to the tangential tension,
respectively.


As derived in \cite{Sakellariadou,traveling_wave-Vachaspati},
$\Sigma$ and $V^{a}$ are induced by the motion of an infinite
Nambu-Goto string within the first order of oscillation
amplitude, while the others are induced in the higher order.  
Their results show that the energy momentum perturbations 
$s^{a}_{\;\;b}$, $s^{a}_{(o1)}$, $s_{(o2)}$ and $s_{(e2)}$ are 
irrelevant to the perturbation of an infinite Nambu-Goto string
within the first order.  
This corresponds to the fact that a Nambu-Goto string is only
charactorized by its energy density equal to its tangential
tension and does not have any other properties such as the
tension normal to its worldsheet, spin or boost along itself.
Hence, in this paper, we concentrate only on $\Sigma$ and
$V^{a}$ and drop the other coefficients in the perturbative
energy momentum tensor (\ref{e-m_decomp_def_1}) and
(\ref{e-m_decomp_def_2}), since we consider the dynamics of an
infinite Nambu-Goto string within the first order of its
oscillation amplitude.



In terms of the gauge invariant variables,
we write the perturbed Einstein equations: for odd modes, 
\begin{eqnarray}
  \label{odd_mode_einstein}
  D^{a}F_{a} = 0, \quad
 (\Delta + \kappa^{2})F_{a} - D^{c}D_{a}F_{c} = 0, 
\end{eqnarray}
and for even modes, 
\begin{eqnarray}
  \label{even_mode_einstein-1}
  && (\Delta + \kappa^{2})F_{ab} = {\cal R}F_{ab} + 2 D_{(a}V_{b)} -
  \gamma_{ab}D_{c}V^{c}, \quad   (\Delta + \kappa^{2})F = 0 , \\
  \label{even_mode_einstein-2}
  && D^{c}F_{ac} - \frac{1}{2}D_{a}F = V_{a}, \quad F^{c}_{\;\; c} = 0,
\end{eqnarray}
where $\Delta := D^{a}D_{a}$.
The perturbative divergence of the energy momentum tensor are
reduced to
\begin{equation}
  \label{even_cnt_eq_of_pem}
  \kappa^{2} V^{a} + \frac{1}{2} {\cal R} D^{a}F  = 0, \quad
  D_{a}V^{a} + \frac{1}{2}\Sigma = 0.
\end{equation}
The first equation corresponds to the Euler equation which
coincides with the equation of the perturbative string motion
derived from the Nambu-Goto action\cite{kouchan-k=0} and the
second equation corresponds to the continuity equation for the
energy density.


We find that
(\ref{odd_mode_einstein})-(\ref{even_cnt_eq_of_pem}) for even
and odd modes on ${\cal M}_{in}$ are reduced to the wave
equations for two scalar variables $\Phi_{(o)}$ and $\Phi_{(e)}$
\begin{equation}
  \label{master-eq}
  (\Delta + \kappa^{2})\Phi_{(o),(e)} = 0,
\end{equation}
respectively, and all gauge invariant variables are given by
$\Phi_{(o)}$ and $\Phi_{(e)}$ without loss of generality as
follows\cite{kouchan-k=0}:
\begin{eqnarray}
  \label{Fa-master}
  &&F_{a} = \epsilon_{ab}D^{b}\Phi_{(o)}, \\
  \label{Fab-F-master}
  && F_{ab} = \left(D_{a}D_{b} - \frac{1}{2}\gamma_{ab}\Delta
  \right) \Phi_{(e)}, \quad F = \Delta\Phi_{(e)}, \\
  \label{Va-Sigma-master}
  && V_{a} = \frac{1}{2}{\cal R}D_{a}\Phi_{(e)}, 
  \quad \Sigma = - {\cal R}\Delta\Phi_{(e)},
\end{eqnarray}
where $\epsilon^{ab}$ is the two-dimensional antisymmetric
tensor on ${\cal M}_{1}$. 
On ${\cal M}_{ex}$, we also find that
(\ref{odd_mode_einstein})-(\ref{even_mode_einstein-2}) are
reduced to the same form as
(\ref{master-eq})-(\ref{Va-Sigma-master}) with ${\cal R} = 0$. 
The exterior solution $\Phi_{(o),(e)}^{(ex)}$ and the interior
solution $\Phi_{(o),(e)}^{(in)}$ to (\ref{master-eq}) are 
\begin{eqnarray}
  \label{master_sol_vac}
  && \Phi_{(o),(e)}^{(ex)} = \sum_{m=0}^{\infty} e^{im\phi} 
  \left\{A H^{(1)}_{\mu}(\beta r) 
    + B H^{(2)}_{\mu}(\beta r) \right\}, \\
  \label{master_sol_homo}
  &&\Phi_{(o),(e)}^{(in)} = \sum_{m=0}^{\infty} e^{im\phi} 
  \left\{C P^{m}_{\nu}(x) + D Q^{m}_{\nu}(x)\right\},
\end{eqnarray}
where $H^{(1)}_{\mu}(\beta r)$ and $H^{(2)}_{\mu}(\beta r)$ are
the Hankel function of the first and the second class, 
$P^{m}_{\nu}(x)$ and $Q^{m}_{\nu}(x)$ are the associated
Legendre function of the first and second class, and 
$ \mu :=  m/(1-\alpha)$, $\beta := \kappa/(1 - \alpha)$, 
$\nu(\nu+1) := \kappa^{2}/\hat\alpha^{2}$ and $x := \sqrt{1 -
\hat{\alpha}^{2}r^{2}}$.
The coefficients $A$ and $B$ correspond to the amplitude 
of the outgoing and the incident wave, respectively.
The regularity condition at the axis $r = 0$ on
$\Phi_{(o),(e)}^{(in)}$ leads $D=0$.



Now, we construct the global solutions to the perturbed Einstein
equations in ${\cal M}$ by matching the exterior and the
interior solutions (\ref{master_sol_vac}) and (\ref{master_sol_homo}) 
along the thick string surface ${\cal S}$ ($r=r_{*}$).
The perturbed solutions should satisfy the perturbed
junction conditions $[\delta q_{\mu\nu}] = 0$ and 
$[\delta K^{\mu}_{\;\;\nu}] = 0$, where $\delta q_{\mu\nu\pm}$
is the perturbed intrinsic metric and 
$\delta K^{\mu}_{\;\;\nu\pm}$ is the perturbed extrinsic curvature
of ${\cal S}$. 
These perturbed quantities are described by the metric
perturbations, which are represented by $\Phi_{(e),(o)}$ through 
(\ref{Fa-master}) and (\ref{Fab-F-master}).
After some calculations, the perturbed junction conditions tell us
\begin{equation}
  \label{junction}
  [\Phi_{(o,e)}] =0, \quad [D_{\perp}\Phi_{(o,e)}] = 0, 
\end{equation}
where $D_{\perp} = n^{a}D_{a}$ and $n^{a} =
(\partial/\partial\rho)^{a}$.

\begin{figure}[htbp]
  \begin{center}
    \leavevmode
    \epsfxsize=8cm
    \epsfbox[0 0 295 242]{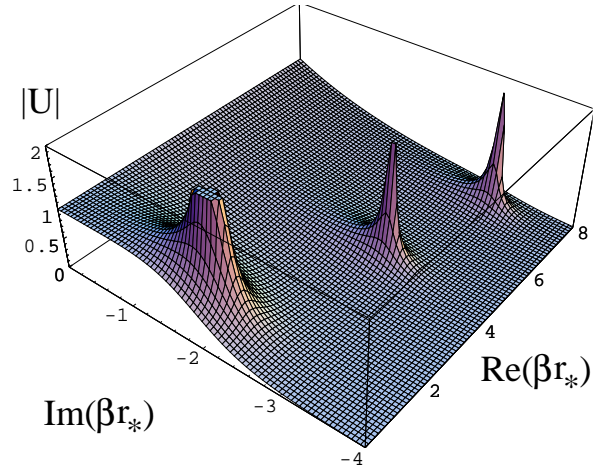}
    \caption{The absolute value of the scattering data $U$ with
    $\alpha=0.3$ and $m=1$ in the complex $\beta r_{*}$ plane is
    illustrated. There are some resonance poles associated with
    the string thickness.} 
    \label{fig:fig2}
  \end{center}
\end{figure}

Substituting (\ref{master_sol_vac}) and (\ref{master_sol_homo}) 
into (\ref{junction}), we have
\begin{eqnarray}
  \label{s_matrix_def}
  A &=& U B, \\
  U &=& - \frac{\kappa r_{*}H_{\mu+1}^{(2)}(\beta r_{*}) 
    P^{m}_{\nu}(x_{*}) - \left(\sqrt{1 - x^{2}_{*}}
    P^{m+1}_{\nu}(x_{*}) + \alpha m P^{m}_{\nu}(x_{*}) \right) 
    H^{(2)}_{\mu}(\beta r_{*})}{\kappa r_{*}H_{\mu+1}^{(1)}(\beta r_{*})  
    P^{m}_{\nu}(x_{*}) - \left(\sqrt{1 - x^{2}_{*}}
    P^{m+1}_{\nu}(x_{*}) + \alpha m P^{m}_{\nu}(x_{*})
    \right) H^{(1)}_{\mu}(\beta r_{*})},
  \label{s_matrix}
\end{eqnarray}
where $x_* := \sqrt{1-\hat\alpha^2 r_*^2} = 1 - \alpha.$
The absolute value of $U$ with $m=1$ and $\alpha=0.3$ is
illustrated in Fig.\ref{fig:fig2}.


The deformation of ${\cal S}$ is represented by $V^{a}$ on
${\cal S}$. 
By the appropriate choice of the function $\zeta^{a}$, we fix
the gauge freedom $X^{a}\rightarrow X^{a} - \zeta^{a}$ in the
neighborhood of ${\cal S}$ so that 
\begin{equation}
  \label{string_displace}
  \left.X_{a} \right|_{{\cal S}}
  := X_{a\pm} 
  = \left. - \frac{1}{{\cal R}}V_{a}\right|_{{\cal S}-} 
  = \left. - \frac{1}{2}D_{a}\Phi_{(e)} \right|_{{\cal S}}. 
\end{equation}
Since $i\int V_{a}\omega S$ is a precise momentum perturbation,
$\int \left.X_{a}\right|_{{\cal S}} S$ does
represent the deformation of ${\cal S}$.


Now, we consider the thin string case.
Physically, a ``thin string'' means a string whose thickness
$r_{*}$ is sufficiently smaller than the wavelength of
gravitational wave.
In this paper, we consider the situation 
$\epsilon := \beta r_{*} \ll 1$ with the finite outside deficit
angle $\alpha$ and take the leading order of $\epsilon$ for
the thin string case.


Further, we note that only $m=1$ mode in (\ref{master_sol_vac})
and (\ref{master_sol_homo}) shows the motion of a Nambu-Goto
thin string.   
$m=0$ and $m>1$ modes are irrelevant to a thin
string\cite{kouchan-k=0}. 
Hence, in the thin string case, the displacement $X_{S}^{a}$ of
a Nambu-Goto string by the gravitational wave scattering is given
by 
\begin{equation}
  \label{string_deformation}
  X^{a}_{S} := \int \left.X^{a}\right|_{{\cal S}}(m=1,\epsilon\ll 1) S = \int
  \frac{\kappa B_{m=1}S}{2 (1-\alpha)\Gamma(\frac{2-\alpha}{1-\alpha})} 
   \left(\frac{\epsilon}{2}\right)^{\frac{\alpha}{1-\alpha}}
   e^{i\phi} \left( n^{a} + i \tau^{a} \right),
\end{equation}
where $\tau^{a} = (1/r) (\partial/\partial \phi)^{a}$ and 
$n^{a} = (1-\alpha) (\partial/\partial r)^{a}$.
(\ref{string_deformation}) shows that the string is deformed
while the incident wave exists on the string. 
In the same order calculation where $X^{a}_{S}$ is given by
(\ref{string_deformation}), we obtain the trivial scattering
data $U\sim 1$ from (\ref{s_matrix}).


The order of magnitude of
$|X^{a}_{S}|:=\sqrt{\gamma_{ab}X^{a}_{S}X^{b}_{S}}$ is estimated
as follows: 
\begin{equation}
  |X^{a}_{S}|/r_{*} \sim
  \kappa^{2}|B_{m=1}|\epsilon^{\alpha/(1-\alpha)}/(\kappa r_{*}) 
  \sim \kappa^{2}|B_{m=1}| \epsilon^{\alpha/(1-\alpha)-1}.
\end{equation}
The amplitude of gravitational wave ($F$ or $F_{ab}$) is
proportional to $|\kappa^2 B| \ll 1$.
If $\epsilon^{1-\alpha/(1-\alpha)}< \kappa^{2} |B|\ll 1$, 
then $|X^{a}_{S}|>r_{*}$. Therefore, the magnitude of the
displacement may become larger than the string thickness within
the linear perturbation framework.


Thus, we have obtain the result that there is neither resonance
nor phase shift in the scattering problem, nevertheless the
string is deformed by the gravitational waves.
In particular, the fact that there is no resonance means an infinite
thin string does not emit gravitational wave spontaneously by
its oscillatory motion.
It should be noted that the trivial scattering does not dictate
the absence of gravitational lensing effect by the deficit angle
$\alpha$ on ${\cal M}_{ex}$.  
We will explicitly see the lensing effect by the scattering of
wave packet which is suitably constructed by
(\ref{master_sol_vac}) because the mode functions already
include the effect of $\alpha$.


Using the linearized Einstein equation, we have found that the
perturbative string displacement $X^{a}_{S}$ is represented by
the gravitational wave $\Phi_{(e)}$. 
In this sense, the dynamical degree of freedom of
the string displacement is given by that of the gravitational
waves on the string surface.
Further, (\ref{string_displace}) and the trivial scattering data
show that the string is bent 
while
the incident wave is passing through the string worldsheet.
These behaviors are same as that for a self-gravitating
spherical Nambu-Goto wall in the first order with respect to its
oscillation amplitude
\cite{Kodama}.
This is our main conclusion.



To obtain the above results, we have first considered the
scattering by a thick string.
We regard that a ``thin string'' is not a string with the
thickness $r_{*}\rightarrow 0$ but that whose thickness $r_{*}$
is sufficiently smaller than the wavelength of gravitational wave. 
For a fixed amplitude of the incident wave, the magnitude of the
string displacement $X^{a}_{S}$ depends on $r_{*}$.
If we take the limit $r_{*}\rightarrow 0$ for fixed the wavelength
of gravitational wave, these is no response of string motion for
finite incident gravitational wave.
This result is consistent with that obtained by Unruh
et.al.\cite{Unruh}. 
Both their and our results show that the straight string cannot
bend in the limit $r_{*}\rightarrow 0$.
This mathematical limit will be irrelevant for strings
formed during phase transition in the early universe, because
they have finite thickness.  
Further, the scattering data (\ref{s_matrix}) has the resonance
poles at $\beta r_{*}\sim 1$ or larger as seen in Fig.\ref{fig:fig2}.
This suggest that a cosmic string oscillates spontaneously and
emits the gravitational waves with the frequencies of the order
of the string thickness. 
In this situation, the dynamics of strings is no longer
approximated by that of thin Nambu-Goto strings.


One might think that our result depends sensitively on the
distribution of $\sigma$ in (\ref{ene_mon}) and our model would
be too artificial because of the step function distribution of
$\sigma$.
Further, the above analysis does not include
$\kappa=0$ mode which includes ``cosmic string traveling waves''
discussed in Ref.\cite{traveling_wave-Vachaspati,traveling_wave-other}, 
and one might think that the dynamical degree of freedom of the
string spontaneous oscillation is in this $\kappa=0$ mode.
However, we obtain the same conclusion in the thin string case
even when the background $\sigma$ is different from the step
function and our conclusion is unchanged in the thin string case 
even if we include the $\kappa=0$ mode into our consideration. 
These two points will be discussed in a separated paper
\cite{kouchan-k=0}.



The authors thank Professor Minoru Omote and Professor Akio Hosoya
for their continuous encouragement.
This work was partially supported by Soryushi Shogakukai and
Yukawa Shogakukai (A.I.).


\end{document}